# USING TWITTER TO PREDICT SALES: A CASE STUDY

*Remco Dijkman[1], Panagiotis Ipeirotis[2], Freek Aertsen[3], Roy van Helden[4]*

[1] *Eindhoven University of Technology, Eindhoven, The Netherlands*

*r.m.dijkman@tue.nl*

[2] *New York University, Stern School of Business, USA*

*panos@stern.nyu.edu*

[3] *EyeOn, Aarle-Rixtel, The Netherlands*

*Freek.Aertsen@eyeon.nl*

[4] *royvanhelden2@hotmail.com*

## Abstract

*This paper studies the relation between activity on Twitter and sales. While research exists into the relation between Tweets and movie and book sales, this paper shows that the same relations do not hold for products that receive less attention on social media. For such products, classification of Tweets is far more important to determine a relation. Also, for such products advanced statistical relations, in addition to correlation, are required to relate Twitter activity and sales. In a case study that involves Tweets and sales from a company in four countries, the paper shows how, by classifying Tweets, such relations can be identified. In particular, the paper shows evidence that positive Tweets by persons (as opposed to companies) can be used to forecast sales and that peaks in positive Tweets by persons are strongly related to an increase in sales. These results can be used to improve sales forecasts and to increase sales in marketing campaigns.*

*Keywords: Sales prediction; Forecasting; Sentiment analysis; Classification; Event study; Twitter.*

# 1    Introduction

Increasingly, consumers post their opinions on social media, commenting on their experiences with products and services they purchased. Researchers and practitioners are also becoming aware that these posts may be correlated to or even cause product success or failure. We have already evidence that there are correlations between social media activity and, for example, movie sales (Asur and Huberman 2010) and book sales (Gruhl 2005). However, we do not have direct evidence about the effect of social media activity on other products, that are less socially-oriented than movies and books.

This paper aims to fill this gap, by studying the relation between social media activity and company sales, for products that create far less activity on social media than movies and books. Focusing on activity on Twitter, the paper shows that, while for movies and books the number of Tweets is directly correlated to sales, this does not hold for other products. As a consequence, more advanced techniques must be used to determine the relation between activity on social media and sales. For that reason, the paper presents a classification of Tweets that is useful when relating Tweets to sales. It then studies the relation between certain classes of Tweets and sales in three ways:
1. By determining whether there is a correlation between the number of Tweets of a certain type and sales in preceding or following weeks.
2. By determining to what extent the number of Tweets of a certain type can be used to forecast sales.
3. By determining whether a high number of Tweets leads to an increase in sales in the following weeks.

Relations of the second and third type are useful in practice. Relations of the second type are useful, because, those relations can be used to improve sales forecasts and therewith production planning. Relations of the third type are useful, because those relations can be used to boost sales in marketing campaigns.

By investigating these relations, the paper has the following contributions. First and foremost, the paper presents a thorough statistical analysis of the relations between Tweets and sales, for products that are less socially-oriented than movies and books. To the best of our knowledge, this relation has not been studied before. The paper shows that, in addition to a classification of Tweets, different types of statistical analysis are required for such products. As a by-product, the paper presents a classification of Tweets that is accurate and effective in the context of making (sales) predictions for a company. While other classifications of Tweets exist (Java et al. 2007; Jansen et al. 2009; Naaman, Boase, and Lai 2010; Sriram et al. 2010; Dann 2010), these classifications are aimed at classifying Tweets in general, or at detecting specific types of Tweets, such as spam (Wang 2010; Benevenuto et al. 2010). They are not targeted at finding the best types of Tweets for making predictions for a company.

Against this background. The remainder of the paper is structured as follows. First, it presents the dataset that will be used to investigate the relation between Twitter activity and sales. Next, it presents a classification of Tweets that is effective when investigating this relation. Then, it investigates the actual relation, using three techniques: by determining the correlation, by determining the possible use of Twitter activity when creating a sales forecast, and by determining the relation between a peak in Twitter activity and sales. Next, it discussed related work on predicting phenomena with Twitter. Finally, it presents the conclusions.

# 2    Data

We use the Tweets and sales of an internationally operating printing company to investigate the relation between Tweets and sales. Due to the restrictions that Twitter places on the number of (historical) Tweets that can be retrieved per hour, we cannot feasibly retrieve all Tweets about the company.

Therefore, we do the analysis per country. Another reason for doing the analysis per country, is that the sales of the company are identified per country, such that we can perform multiple analyses for the company; one per country. We focus on countries in Europe (Germany, France, Spain, and the Netherlands), because the use of different languages in those countries, combined with the time zone settings of Twitter users, makes it easy to distinguish the country from which a Tweet originates. Note that, while countries may be in the same time zone, Twitter time zones are identified by the capital of a country, such that confusion between countries in the same time zone in which the same language is spoken (such as Germany, Austria, and Switzerland) should be rare. We use data of 2012 and the first three quarters of 2013. Table 1 shows the number of Tweets about the company per country during that timeframe. The Tweets were collected by using the Twitter API to search for the company name, in combination with the language and time-zone of the country of origin. The table shows that there is a striking difference in the number of Tweets per country, which we mainly attribute to the difference in use of social media in different countries (Takhteyev, Gruzd, and Wellman 2012).

| **Country** | **Nr. of Tweets** | **Population** |
|---|---|---|
| France | 1,791 | 63 million |
| Germany | 3,218 | 81 million |
| Spain | 4,062 | 47 million |
| The Netherlands | 3,709 | 17 million |
| **OVERALL** | 12,780 | |

**Table 1. Number of Tweets per Country.**

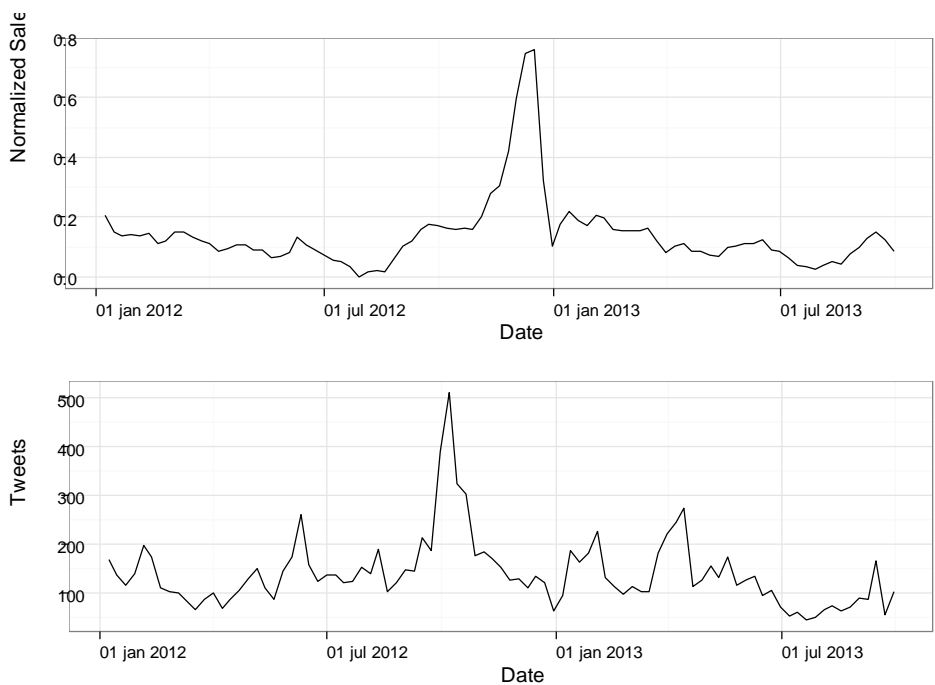

**Figure 1. Sales and Tweets per week.**

Figure 1 shows the total sales and tweets per week for the four countries combined. For confidentiality the sales have been normalized. We aggregate the sales per week and not per day, because there is a strong weekly pattern in the data, with low sales during the weekend and a strong peak on Monday. Aggregating the sales and Tweets per week eliminates accidental effects due to that pattern.

The graphs show a fairly stable distribution of the sales, with a mild dip during the summer and a strong peak at the end of the year. The Tweets are far less stable. There are some strong peaks in the data. A visual inspection of the Tweets provides explanations for some of the peaks. The peak at half of June 2012 is caused by both German and Dutch Tweets about a legal dispute of the company. The strong peak in October 2012 is caused by a high number of spam messages in Spain. The peaks in January 2013 are caused by Spanish Tweets that point out a URL where discount vouchers can be downloaded. The peaks in April and September 2013 are caused by German and Dutch spam messages, respectively. As spam may influence buyer behaviour, we deliberately keep the spam messages.

For every Tweet, we collect not only the Tweet itself, but also meta-data, in particular: the name, screen name, and number of followers of the user who created the Tweet, as well as the number of retweets of the Tweet. We use this information at a later stage to classify the Tweet and to determine the impact of a Tweet and, therewith, the likelihood that Tweets led to an increase in sales.

## 3    Classification

We classify Tweets, because we hypothesize that different types of Tweets have a different effect on sales. For example, negative Tweets can be assumed to have a negative effect on sales, while positive Tweets can be assumed to have a positive effect on sales. Classifying Tweets allows us to explore the different effects of different types of Tweets. We classify Tweets in three dimensions: the type of Tweet, the type of user who posted the Tweet, and the sentiment of the Tweet.

There exist a number of classifications of types of Tweets (Java et al. 2007; Jansen et al. 2009; Naaman, Boase, and Lai 2010; Sriram et al. 2010; Dann 2010). However, these classifications primarily focus on personal communications and, as a consequence, have many different classes for different types of personal communication. We focus on communication about a company or product and consequently require a more fine-grained classification for such messages. Looking both at existing classifications and at the Tweets that we collected, we defined the following types of Tweets:

- Job advert
- Product advert
- Positive or negative customer experience report
- Response to a customer experience report
- Daily chatter
- Factual statements about something that was bought
- Requesting information about the company or its products
- Pointing out, or providing information or advise about the company or its products
- News broadcast about the company
- Other

Some of these classes can be expected to overlap and indeed do overlap as we will see further on in the paper. Therefore, we aim to adapt the classification by combining classes after the data collection for the classification.

For the classification of type of user, we use the classes 'person', 'the company of the study', and 'other organizations'. For the classification of sentiment, we use the classes 'positive', 'neutral', and 'negative'.

We first perform a manual classification. 171 people performed the manual classification. Each Tweet was classified by three different people. In total 2,511 Tweets were classified by three people. We use a Tweet's classification, in case it received the same classification by at least two people. Accordingly, we define the accuracy of a Tweet's classification as follows: the classification of a Tweet is as a 'hit' when somebody else chose the same classification; the classification of a Tweet is a 'miss' when nobody else chose the same classification.

| Type of Tweet | Acc. |
|---|---|
| Job advert | 92% |
| Product advert | 82% |
| Customer experience | 72% |
| Response to experience | 50% |
| Chatter | 50% |
| What was bought | 78% |
| Information request | 59% |
| Advice | 54% |
| News | 40% |
| Other | 26% |
| **OVERALL** | **69%** |

| Type of User | Acc. |
|---|---|
| Person | 89% |
| Company | 43% |
| Other organizations | 88% |
| **OVERALL** | **85%** |

| Sentiment | Acc. |
|---|---|
| Positive | 84% |
| Neutral | 74% |
| Negative | 84% |
| **OVERALL** | **80%** |

**Table 2. Accuracy of Manual Classifications.**

Table 2 shows the accuracy of the manual classification per class. It shows particularly low accuracy in the classification of the type of tweet and in the classification of the type of user, when the type of user is the company itself. For that reason, we aggregate some of the classes. In particular, we aggregate the 'news' and 'other' class as well as all the classes that relate to personal communication ('customer experience', 'response to customer experience', 'chatter', 'what was bought', 'information request', and 'advise'). We also aggregate 'company' and 'other organizations', and we aggregate 'neutral' and 'negative'. The resulting classification and accuracy is shown in Table 3. As can be seen, this strongly increases the accuracy. While the 'other' category is still rather inaccurate, it only represents a small part (5%) of the total number of classifications and, therefore, does not decrease the overall accuracy much.

| Type of Tweet | Acc. |
|---|---|
| Job advert | 92% |
| Product advert | 82% |
| Personal messages | 90% |
| Other | 39% |
| **OVERALL** | **85%** |

| Type of User | Acc. |
|---|---|
| Person | 89% |
| Organization | 91% |
| **OVERALL** | **90%** |

| Sentiment | Acc. |
|---|---|
| Positive | 84% |
| Not positive | 86% |
| **OVERALL** | **85%** |

**Table 3. Accuracy of the Revised Manual Classification.**

We use the manually classified Tweets as a training and test set to automatically classify all 12,780 Tweets that were collected. We aim to perform the classification in a language-independent manner. Therefore, we do not use text classification. Instead, we rely on properties of Tweets that can be measured independent of the language. In particular, we record for each Tweet:
- the number of retweets
- whether the Tweet is a retweet itself
- the number of hyperlinks contained in the Tweet
- the number of hashtags contained in the Tweet
- the number of user references contained in the Tweet

- whether the Tweet contains emoticons
- the number of question marks contained in the Tweet
- the number of exclamation marks contained in the Tweet
- the number of followers of the user that posted the Tweet
- the number of friends of the user that posted the Tweet
- the total number of Tweets of the user that posted the Tweet
- whether the username of the user that posted the Tweet contains a common first-name

We use a decision tree to perform the classifications, because this allows us to inspect the rules that the classification is based on. We use an 80% training set and a 20% test set. Table 4 shows the accuracy of the automatic classification on the test set. The table both shows the overall accuracy and the accuracy per language. We did not have any accurately manually classified French Tweets. As a consequence, we cannot determine the accuracy of French Tweets.

| Country | Accuracy Type of Tweet | Accuracy Type of User | Accuracy Sentiment |
|---|---|---|---|
| France | --- | --- | --- |
| Germany | 84% | 96% | 81% |
| Spain | 75% | 93% | 77% |
| The Netherlands | 87% | 84% | 75% |
| **OVERALL** | **86%** | **86%** | **76%** |

**Table 4. Accuracy of the Automatic Classification.**

We attribute the differences in accuracy between the different countries and languages mainly to differences in the properties of the training and test sets. For Germany, the large majority of the Tweets are advertisements, which is relatively easy to classify for a single company, because it is strongly repetitive. The Spanish and Dutch set is more diverse, but the Dutch set contains a relatively large number of job advertisements, which are again relatively easy to classify.

Looking at the decision trees that are produced, not surprisingly, the presence of a first name in the user name is a very strong indicator of whether a user is a person or an organization. Also taking into account the number of hyperlinks in the Tweet (a hyperlink typically indicates a company) already leads to an accuracy of 83%. Rules for determining the type of a Tweet are harder to define. However, a high number of Tweets by the user indicates an advertisement in 84% of the cases, which can either be a job advertisement or a product advertisement. This corresponds to the idea of spam. Spammers typically send a high number of messages. When the number of Tweets is relatively low and the number of friends of the user is relatively low, we are again looking at a probable advertisement. Taking these rules into account leads to an accuracy of 77%. Interestingly, the first rule for determining whether or not a Tweet is positive is not related to the content: a high number of Tweets sent by the user indicates a positive sentiment. Looking at this rule closely, it is not a surprising conclusion, because spam would be considered to generally carry a positive message (which is not the same as saying that it is perceived positively by the recipient) and spam is associated with a high number of Tweets by the user as well. The second rule to take into account is the number of exclamation marks in the Tweet: the presence of an exclamation mark typically indicates a positive sentiment. Taking these two rules into account leads to an overall accuracy of 73%.

We use the classified Tweets to investigate the effect of different types of Tweets on sales.

# 4 Relation

We investigate the relation between Tweets and sales in three ways:
1. By determining the correlation between the number of Tweets in a week and the number of sales in the following weeks.
2. By determining whether Tweets can be used to improve sales forecasts.
3. By determining whether a high number of Tweets in a week leads to a significantly higher number of sales in the following weeks.

## 4.1 Correlation between sales and Tweets

The first step in determining the relation between Tweets and sales is determining the correlation between the number of Tweets of a certain type in a week and the sales in subsequent weeks. The reason for this is that, if there is a correlation, that would be a first quantitative proof that Tweets of a certain type can be used to predict or even influence sales. If a correlation can be found, this is ground for further exploring this particular correlation.

We investigate the correlation per country and per type of Tweet, type of user and sentiment. We look at the correlation with sales one to four weeks into the future. We also look at the sales one to four weeks in the past, to check the direction of the correlation; it could be that sales follow Tweets, but it could also be that Tweets follow sales. The latter can easily be explained: people buy something and when they receive it, they Tweet about it. However, it cannot be used to predict or influence sales.

Table 5 shows the correlations between sales and Tweets for Spain and the Netherlands, using Pearson's R. (Moderate) correlations are shown in boldface. We did the same analysis for France and Germany, but did not find any correlations for those countries. Therefore, we did not include tables for those countries. Each row in the table, shows the correlations for that particular type of Tweet, where the type of Tweet is identified as: sent by a per(son) or org(anization), a Tweet about a product ad(vertisement) or p(ersonal) c(ommunication), and whether or not a Tweet is pos(itive). In this way, the row marked 'per' 'all' 'pos', for example, shows all correlations of Tweets by persons with a positive sentiment. The table does not include job advertisements as a class, because we do not expect job advertisements to be causally related to sales and, therefore, do not consider them interesting to investigate. The table also excludes personal communication as a subclass of Tweets by companies. This is because personal communication is defined as communication between people and therefore, by definition, excludes Tweets by organizations. The 'other' class is also excluded, because correlations with Tweets in that class will be impossible to explain. The columns show the correlations from 4 weeks before the week for which we measure the Tweets until 4 weeks after.

The tables show that there is a moderate correlation, both for Spain and for the Netherlands, between positive Tweets by persons and the sales in the third and fourth week following the Tweets. For the Netherlands there is also a moderate correlation in the second week following the Tweets. All correlations are significant at $p<0.01$. The tables show that there already is a correlation between sales and positive Tweets in general, but when further restricting the Tweets to positive Tweets by *persons*, the correlation becomes stronger. Interestingly, this does not hold for organizations; positive Tweets by organizations are not correlated to sales.

Based on these results, we can say that positive Tweets by people are positively related to sales in the weeks following the Tweets. This hypothesis has a good basis, because it is found in two countries independently. Also, there is a logical explanation for the correlation: when people say something positive about an organization, this causes other people to buy from that organization.

While this result is promising, it is unclear how it can be exploited. We look for exploitations in two directions. Firstly, we investigate whether the correlation between Tweets and sales can be used to improve sales forecasts. Secondly, we investigate whether the correlation can be used to influence sales positively.

| User | Tweet | Sent | Correlation to sales in week (Spain) | | | | | | | | | Correlation to sales in week (Netherlands) | | | | | | | | |
|---|---|---|---|---|---|---|---|---|---|---|---|---|---|---|---|---|---|---|---|---|
| | | | -4 | -3 | -2 | -1 | 0 | 1 | 2 | 3 | 4 | -4 | -3 | -2 | -1 | 0 | 1 | 2 | 3 | 4 |
| all | all | all | 0.06 | 0.07 | 0.08 | 0.10 | 0.09 | 0.09 | 0.13 | 0.18 | 0.23 | -0.08 | -0.03 | 0.05 | 0.15 | 0.23 | 0.27 | 0.25 | 0.21 | 0.21 |
| | all | pos | 0.17 | 0.13 | 0.15 | 0.16 | 0.10 | 0.11 | 0.23 | **0.30** | **0.33** | 0.00 | -0.02 | -0.02 | 0.03 | 0.15 | 0.25 | **0.32** | **0.36** | **0.35** |
| | ad | all | -0.02 | 0.01 | -0.01 | 0.03 | -0.01 | -0.02 | 0.05 | 0.03 | 0.01 | -0.08 | -0.08 | -0.06 | -0.01 | 0.07 | 0.18 | 0.24 | 0.25 | 0.22 |
| | pc | all | 0.06 | 0.07 | 0.08 | 0.10 | 0.09 | 0.09 | 0.13 | 0.18 | 0.23 | -0.04 | 0.02 | 0.11 | 0.20 | 0.26 | 0.22 | 0.14 | 0.08 | 0.10 |
| | pc | pos | 0.20 | 0.14 | 0.16 | 0.16 | 0.11 | 0.14 | 0.23 | **0.32** | **0.36** | 0.05 | 0.09 | 0.12 | 0.09 | 0.12 | 0.15 | 0.14 | 0.20 | 0.17 |
| per | all | all | 0.14 | 0.15 | 0.16 | 0.18 | 0.16 | 0.14 | 0.16 | 0.18 | 0.22 | -0.06 | -0.02 | 0.06 | 0.14 | 0.22 | 0.27 | 0.26 | 0.21 | 0.19 |
| | all | pos | 0.12 | 0.15 | 0.17 | 0.16 | 0.10 | 0.12 | 0.24 | **0.32** | **0.34** | 0.02 | -0.01 | -0.01 | 0.04 | 0.15 | 0.26 | **0.34** | **0.38** | **0.36** |
| | ad | all | -0.02 | 0.01 | -0.01 | 0.03 | -0.01 | -0.02 | 0.05 | 0.03 | 0.01 | -0.08 | -0.08 | -0.06 | -0.01 | 0.07 | 0.18 | 0.24 | 0.25 | 0.22 |
| | pc | all | 0.14 | 0.15 | 0.16 | 0.18 | 0.16 | 0.14 | 0.16 | 0.18 | 0.22 | -0.01 | 0.04 | 0.13 | 0.21 | 0.26 | 0.22 | 0.15 | 0.07 | 0.07 |
| | pc | pos | 0.14 | 0.16 | 0.19 | 0.16 | 0.11 | 0.15 | 0.25 | **0.34** | **0.38** | 0.15 | 0.16 | 0.17 | 0.15 | 0.15 | 0.21 | 0.23 | 0.29 | 0.25 |
| org | all | all | -0.11 | -0.10 | -0.08 | -0.06 | -0.05 | -0.02 | 0.06 | 0.14 | 0.22 | -0.12 | -0.06 | -0.01 | 0.11 | 0.16 | 0.11 | 0.08 | 0.09 | 0.18 |
| | all | pos | 0.17 | 0.01 | 0.02 | 0.07 | 0.04 | 0.03 | 0.06 | 0.09 | 0.11 | -0.25 | -0.15 | -0.10 | -0.11 | -0.05 | -0.11 | -0.22 | -0.18 | -0.17 |
| | ad | all | 0.00 | 0.00 | 0.00 | 0.00 | 0.00 | 0.00 | 0.00 | 0.00 | 0.00 | 0.00 | 0.00 | 0.00 | 0.00 | 0.00 | 0.00 | 0.00 | 0.00 | 0.00 |

**Table 5. Correlation between Tweets and Sales in Spain and the Netherlands**

### 4.2 Forecasting sales with Tweets

We determine whether the correlation between Tweets and sales can be used to improve sales forecasts, using a Granger test.

In order to execute the Granger test, we must ensure that the data is stationary. This is not the case. We already showed in Figure 1 that there is a peak at the end of the year and a dip during the summer. An augmented Dickey-Fuller test also shows that neither the Tweets nor the sales are stationary. We make the data stationary, by taking the first order difference of both time series.

We perform the Granger test on the resulting data and four lags of both Tweets and sales. The Granger test reveals no Granger causality between the two time series, neither for the Dutch nor for the Spanish data set.

For the Spanish data set, there is significant (at $p=0.05$) Granger causality with five lags or more. When, rather than measuring the number of positive Tweets by people, we measured the fraction of positive Tweets by people in a week (i.e.: the number of positive Tweets by people in a week divided by the total number of Tweets in a week), we get slightly better results. The fraction of Tweets is stationary, albeit at $p=0.10$, which is not very certain. If we assume that the fraction of Tweets is indeed stationary, there is significant (at $p=0.05$) Granger causality for two or more lags. If we assume that the fraction of Tweets is not stationary, there is significant Granger causality (at $p=0.01$) with five lags or more.

Summarizing, there is some evidence that Tweets can be used to improve sales forecasts. However this evidence only exists for the Spanish dataset, and only when we accept a low significance with respect to stationarity, or when we are making a forecast far (five weeks or more) into the future.

## 4.3   The effect of a high number of Tweets on sales

We investigate the possibility of influencing sales by influencing Tweets, by determining whether a high number of positive Tweets by people leads to an increase in sales. If this is the case, it is possible to influence sales by creating a boost in positive Tweets by people. We could achieve that, for example, by sending vouchers via Twitter, by stressing positive properties of the company, or by asking employees to Tweet positively about the company on their personal account.

We investigate whether a high number of positive Tweets by people leads to an increase in sales, by doing an event study. In the event study, we define a 'high number of positive Tweets by people' as an event, and study the effect of that on sales. What remains is to define what we mean by a 'high number'. At this moment, we define a 'high number' of Tweets as a number of Tweets in the $10^{th}$ 10-quantile (i.e. a number of Tweets in the 10% weeks with the highest Tweets). Clearly, there are other ways of defining what a 'high number' of Tweets is and further on in this section, we will also investigate the robustness of the results for other definitions.

Other important parameters that must be decided on when conducting an event study are the event window and the window for measuring normal sales. There is no strong consensus on the ideal values for either parameter (Konchitchki and O'Leary 2011), although in general the event window is rather short (a couple of days) while the window for measuring normal returns in much longer (tens to hundreds of days). It is important to note, however, that event studies are typically conducted on stock prices rather than sales, such that parameter values do not translate automatically. In particular, we assume that the effect of a commercial on sales lasts longer than the effect on stock price (if the effect of a commercial on stock price is measurable at all). For those reasons, we will again investigate the robustness of the event study for different values of the parameters.

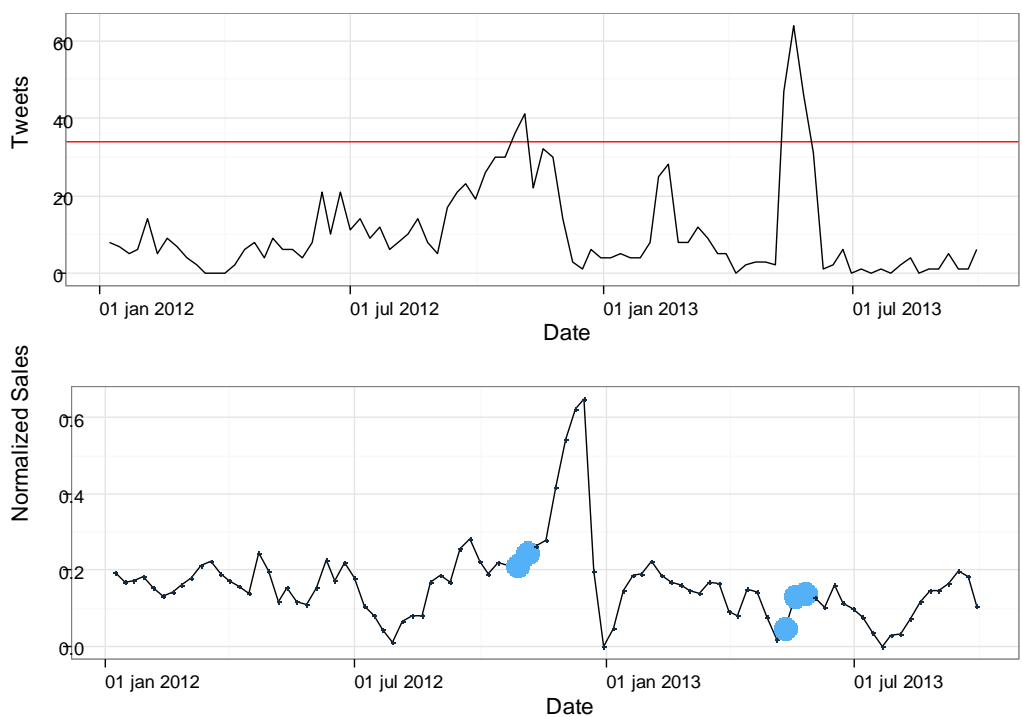

**Figure 2. Dutch sales and Tweets per week with peaks in positive personal Tweets.**

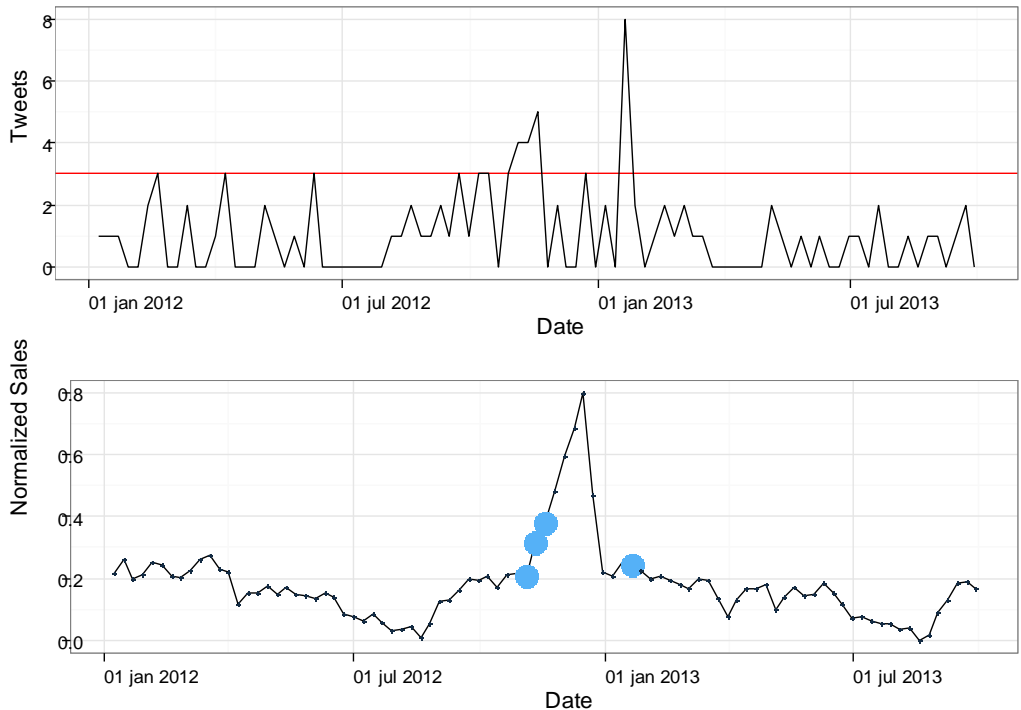

**Figure 3. Spanish sales and Tweets per week with peaks in positive personal Tweets.**

Figure 2 and Figure 3 show the Dutch and Spanish Tweets and sales. The lines in the graphs that show the Tweets indicate the highest 10% quantiles. The weeks that fit in these quantiles are also indicates in the sales graphs with thick points. The graphs already provide information that is relevant to the event study. An important observation is that, while the Spanish set contained more Tweets than the Dutch set, it contains less positive Tweets by people. In fact, the Spanish set contains very few positive messages by people; on average less than 2 per week. There is no clear explanation for this. It seems that either the tone of Spanish messages is less positive than that of Dutch messages, or the people who manually classified the Spanish messages were less positive in their classification. Interestingly, both the Dutch and the Spanish Tweets show a bump from September to November 2012. In addition, the Spanish Tweets show a strong peak in January 2013 and the Dutch Tweets show a strong peak in May 2013. There is no clear explanation for the bumps in September to November 2012. These consist of (positive personal) Tweets with widely varying content. The peak in Spanish Tweets in January 2013 was caused by users pointing out places to each other where coupons could be found. The peak in Dutch Tweets in May 2013 is primarily caused by a single person sending product advertisements. Finally, the peaks in positive personal Twitter traffic indeed seem to precede peaks in sales.

The event study provides statistical proof of this phenomenon. Even though there is only a small number of 'events': four for the Spanish case and five for the Dutch case, a statistically significant increase in Tweets can be detected. For the Dutch case, this increase is robust, using an event window between 0 and 5 weeks after the event, a window for measuring normal sales of up to 10 weeks before the event, and an interpretation of 'high number of Tweets' in the 5% to 20% quantile. The Spanish case is less robust. It is robust using an event window between 0 and 3 weeks after the event and an interpretation of 'high number of Tweets' in the 5% to 20% quantile. These parameter settings are fairly typical. However, it is only robust for a window for measuring normal sales of up to 3 weeks before the event, which is rather short.

Concluding, we can state that for the Dutch set, a high number (in the 5-20% range of highest numbers) of positive personal Tweets is followed by an increase in sales. For the Spanish set, there is some evidence to suggest that a high number of positive personal Tweets is followed by an increase in sales.

As we are conducting a case study, it is also possible to look into the reason why this statistical relation may exist. Specifically, we can investigate why a specific high level of Tweets might cause an increase in sales. Looking at the graph for the January 2013 event, this particular peak in sales does not appear to cause an increase in sales. A possible explanation for the May 2013 event is that the product advertisements sent during this event indeed caused people to buy, thus increasing sales. However, a more likely explanation is that the sales were rather low at the beginning of May 2013 and simply increased to normal levels, which, purely by coincidence, corresponded to a high number of Tweets. Consequently, no clear conclusions about causality can be drawn from these events. The events in October and November 2012 did not consist of clearly identifiable kinds of Tweets. In addition, these events precede the end-of-year peak in sales. Therefore, the most likely explanation for the relation between these events and the increase in sales that follow them, is that there is a third factor in play, which works as follows. Towards the end of the year, people tend to buy more of the company's products. They already start looking at the products in October and, as a consequence, they start Tweeting about them. Therefore, based on this analysis, we cannot claim that there is a causal relation between the peaks in Tweets and the sales, but rather that there is a causal relation between interest in the products and Tweets and between interest in the products and sales.

Even if the relation between Tweets and sales is not causal, it remains interesting. In particular, because companies are interested in predicting the height of peaks in their workload for different reasons. In particular, because companies are dimensioned to peaks in activity and because companies plan their resources in advance to handle peaks in their workload. Knowing the height of the peaks in Tweet activity may help to predict the height of the sales peak. Our current dataset is too small to measure the extent to which this is possible, but we will investigate this further in future work.

At the same time, there is still evidence that a causal relation exists as well; the correlation between the level of Tweets in general (not just the peaks) and sales has been shown in a previous section. One explanation for such a causal relation is that, during high peaks in Twitter activity, a significantly higher number of potential customers is reached, each of which has a probability of becoming an actual customer. In this way, during high levels of Tweets, there is a higher probability of people becoming customers than during average levels of Tweets. This principle is often exploited and, for example, the reason why research is done towards finding 'influential people' on social media (Cha et al. 2010; Bakshy et al. 2011). Indeed, it is the case that during weeks of peak Tweet activity, a far larger number of people is reached than during weeks with average Tweet activity. For the Dutch dataset, the number of followers that receive positive personal Tweets during a week of peak activity is 27,000 on average, the number of followers that receive positive personal Tweets during a week of average activity is 8,000.

Summarizing, peaks in positive Tweets by people are related to an increase in sales. Evidence as to why this may be the case is inconclusive. There is evidence to support a causal relation. In particular, because during a peak week far more potential customers are reached than during an average week. There is also evidence to support the existence of a third factor. In particular of the existence of the third factor that people show interest in products during a certain period of the year and, therefore, both buy the products and Tweet about them. Possibly, both the causal relation and the third factor play a role. To investigate the possibility of a causal relation, we plan to run an experiment in future work. For this experiment, we aim to create a peak in positive personal Tweets during an average week and measure the effect of this peak on sales.

## 5 Related work

Closest to the work presented in this paper is the work on predicting movie box office sales and book sales. Asur and Huberman (2010) predict movie box office sales using targeted queries to search for Tweets that relate to a specific movie. They then relate those numbers to box office sales. The paper also investigates the effect of classifying the Tweets based on sentiment. Subsequently, it performs a regression analysis to predict box office sales. A similar analysis has been done for book sales, but using blog data rather than Twitter data (Gruhl 2005). The work by Asur and Huberman (2010) and Gruhl (2005) is close to the work presented in this paper, because it also predicts sales. However, far more Tweets are produced about movies and books than about arbitrary companies, making the use of standard regression techniques for predicting sales more feasible. Indeed the initial correlation scores in these papers are much higher than the ones in this paper. Therefore, as we show in this paper, when relying on a smaller number of Tweets to predict sales, other statistical methods must be used. Also, classification techniques for Twitter messages become more important. One cannot solely rely on the number of Tweets about a product (movie or book). Instead, the influence depends on the type of Tweet, the type of user that wrote the Tweet and the sentiment of the Tweet.

In addition to predicting movie and book sales, there exists a growing body of knowledge on predicting large social and economic events, in particular: unemployment rates, influenza epidemics, the stock market, and elections.

Antenucci et al. (2014) predict unemployment rates using Twitter data. They uses targeted queries to search for Tweets that are related to job loss and subsequently perform principal component analysis and regression to predict the unemployment rate.

Culotta (2013) predicts influenza infection rates using Twitter data. Predicting influenza infection rates is a popular research area and has been the topic of many research papers, which not only use Twitter data (St Louis, and Zorlu 2012; Paul, and Dredze 2011; Aramaki, Maskawa, and Morita 2011; Lampos, De Bie, and Cristianini 2010), but also other Internet data sources (Ginsberg et al. 2009). These papers use targeted queries to search for Tweets that relate to influenza infections. One of the main challenges in this area of research is to determine whether or not a Tweet is related to an actual influenza infection. For this, classification techniques are used. Subsequently, regression techniques are used to predict influenza infections. Of particular interest in these papers is that they also attempt to predict the peaks of influenza infection rates. This is of interest in the context of sales, because predicting a sales peak helps companies to improve their production planning.

Bollen, Mao and Zeng (2011) predict the stock market as a whole using Twitter. They rely on a sentiment-based classification of Tweets in general. Similar to our second analysis, they use a Granger test to investigate whether Tweets can be used to improve the forecast of a stock market index, showing that this is indeed possible. Zhang, Fuehres, and Gloor (2011) also predict the stock market. Similar to our first analysis, they use correlations with delay.

Tumasjan et al. (2010) use Twitter to predict the outcome of elections. They show that the fraction of attention that certain parties receive on Twitter corresponds to the outcomes of elections. Similar analyses have been done by Bermingham and Smeaton (2011), and Kim and Hovy (2007). All three papers use forms of Tweet classification, based on both sentiment and pre-defined terms. The work on predicting elections with Twitter has also received significant criticism (Gayo-Avello 2012; Jungherr, Jürgens, and Schoen 2012). The criticism raised in these papers applies to any form of prediction using social media. Gayo-Avello (2012) and Jungherr, Jürgens, and Schoen (2012) raise the methodological concerns that, when using social media, there always is a selection bias, and there is the potential for influencing the results, by actively participating on social media. They also raise the concern that the variables in most studies are not carefully selected in advance, but rather are selected after it has become clear which ones produce the best results.

# 6      Conclusion

This paper investigates the relation that exists between activity on Twitter and company sales. There already exists evidence that there is a relation between activity on Twitter and movie or book sales. However, books and movies can rely on high levels of Twitter activity, while typical products cannot. This paper indeed shows that there is a much smaller initial correlation between Twitter activity and products that generate less Twitter activity. It also shows that, while for books and movies, simply counting the number of Tweets provides enough information to predict sales, this is not true for all products. Instead, for products that generate less Twitter activity, only certain classes of Tweets relate to company sales.

Through a case study, involving four separate datasets from separate countries, this paper shows a moderate correlation between positive Tweets by persons (as opposed to companies) and sales in two of the four countries. This suggests that positive Tweets by persons may be used to predict sales and that such Tweets may actually cause sales, such that they could be influenced to increase sales.

In a further investigation, the paper shows with a Granger test that there is evidence that the number of positive Tweets by persons can be used to predict sales. That evidence, however, only holds for one of the four countries that was studied and only when a low level of significance is accepted with respect to stationarity of the Tweets, or when a long-term prediction (for five or more weeks into the future) is made.

Using an event study, the paper shows a highly significant relation between a peak in the number of positive Tweets by persons and an increase in sales in the following weeks. The relation is shown for two of the four countries. Further investigation reveals that the relation is not necessarily a causal relation, but that there is possibly a third factor in play. The investigation also reveals that the relation may be usable in predicting (the height of) a peak in sales.

Since these relations are shown in a case study, we must answer the question of whether the case study is representative. We argue that this question should be answered from the perspective of Twitter activity: companies and products that receive similar attention on Twitter could expect to see similar relations to the ones found in this paper. When products receive far more attention – as is the case for movies and books – related research shows that the relations between Twitter activity and sales is much stronger. Conversely, if products receive less attention, it can be expected that few or no relations can be found between Twitter activity and sales. This claim is supported by the findings in this paper, because most relations were found for the Spanish dataset, which also contained the most Tweets, while no relations were found for the German and French dataset, which also contained the fewest Tweets. This leads to the careful conclusion that at least around 40 Tweets per week are necessary in order to make (weekly) sales predictions based on Twitter.

In addition to replicating the research in other case studies, we aim to conduct future work in two directions. First, we aim to investigate precisely how Twitter activity can be used in sales forecasting. Of particular interest is the forecasting of peaks in sales, because the evidence that this was possible was strongest, and because sales peaks are important when dimensioning a company (e.g. planning maintenance and hiring temporary workers). Second, we aim to determine whether there is a causal relation between Tweets and sales. To this end, we are planning an experiment, in which we create a peak in positive Tweets by people and test whether that leads to an increase in sales.